\documentstyle[12pt]{article}
\def\appendix#1{
\addtocounter{section}{1}
\setcounter{equation}{0}
\renewcommand{\thesection}{\Alph{section}}
\section*{Appendix \thesection\protect\indent #1}
\addcontentsline{toc}{section}{Appendix \thesection\ \ \ #1}
}

\newcommand{\K}{{\cal K }}
\def\be{\begin{equation}}
\def\la{\label}
\def\ee{\end{equation}}
\def\bea{\begin{eqnarray}}
\def\eea{\end{eqnarray}}
\def\eps{\varepsilon}
\def\vphi{\varphi}
\def\d{\delta}

\def\k{{\bf k}}
\def\x{{\bf x}}
\def\y{{\bf y}}
\def\p{\partial}

\def\ttphi{\bar \phi}
\begin{document}
\title{
{\bf Massive symmetric tensor field on AdS}
}
\author{Aleksey Polishchuk\thanks{alexey@mi.ras.ru}
\mbox{} \\
\vspace{0.4cm}
Steklov Mathematical Institute,
\vspace{-0.5cm} \mbox{} \\
Gubkin str.8, GSP-1, 117966, Moscow, Russia
\vspace{0.5cm} \mbox{}
}
\date {}
\maketitle
\begin{abstract}
The two-point Green function of a local operator in CFT
corresponding to a massive symmetric tensor field on the $AdS$
background is computed in the framework of the $AdS/CFT$ correspondence.
The obtained two-point function is shown to coincide with the 
two-point function of the graviton in the limit when the AdS mass vanishes.
\end{abstract}

\section{Introduction}
The AdS/CFT correspondence conjectured in \cite{Mald}
states that in the large $N$ limit and at large
`t Hooft coupling $\lambda = g_{YM}^2 N$ the classical supergravity/M-theory
on the Anti-de Sitter space (AdS) times a compact manifold is dual to a
certain $SU(N)$ conformal gauge theory ($CFT$) defined on the boundary of
$AdS$. One notable example of this correspondence being the duality between
$D=4$, ${\cal N} = 4$ supersymmetric Yang-Mills theory and $D=10$ Type IIB
supergravity theory on $AdS_5 \times S^5$. The precise formulation of the
conjecture was given in
\cite{GKP}, \cite{W} where it was proposed to identify the generating functional for
connected Green functions of local operators in CFT with the
on-shell value of the supergravity action under the restriction that
the supergravity fields satisfy the Dirichlet conditions on the
boundary of AdS. Recall that in the
standard representation of AdS as an upper-half space $x_0 > 0 $,
$x^k \in {\bf R}$, $k = 1,\ldots d$ with the metric
$$
ds^2 = g_{\mu \nu} dx^{\mu} dx^{\nu} = \frac{1}{x_0^2}
\left( dx^0 dx^0 + dx^i dx^i \right) ,
$$
the boundary includes the plane $x_0 = 0$ as well as the point
$x_0 = \infty$. Since the boundary is located infinitely far away from any
point in the interior the supergravity action is infrared divergent and
must be regularized. As was pointed out in \cite{FMMR} the consistent
regularization procedure (with respect to Ward identities) requires one
to shift the boundary of AdS to the surface in the interior defined by
$x_0 = \eps$. Then the Dirichlet boundary value problem for supergravity
fields is properly defined and one can compute the on-shell value of the
supergravity action as a functional of the boundary fields. With the
account of this regularization procedure the standard formulation of the
AdS/CFT correspondence assumes the form:
$$
\langle {\cal O}(\x_1) \cdots {\cal O}(\x_n) \rangle =
\lim_{\eps \to 0} \frac{\d}{\d \phi_1(\x_1)} \cdots \frac{\d}{\d \phi_n(\x_n)}
S_{on-shell} \left( \phi_1(\x_1), \ldots, \phi_n(\x_n) \right) \bigr|_{
\phi_i(\x_i) = \phi_i(x_0 = \eps , \x_i)} ,
$$
where $\x_i$ are some points on the boundary of $AdS_{d+1}$, 
${\cal O}_i(\x_i)$ are
gauge invariant composite operators in CFT and $\phi_i(\x_i)$ are the
corresponding supergravity fields. Here we used the convention in which
the coordinates $x^{\mu}$ of $AdS_{d+1}$ are split according to:
$x = (x_0,\x)$, so that $\x \in {\bf R}^d$. 
The action
$S_{on-shell}$ is the sum of the bulk supergravity action and the boundary
terms necessary to make the AdS/CFT correspondence complete.
The origin of these boundary terms was elucidated in \cite{AF2} where it was
shown that they appear in passing from the Hamiltonian description of the
bulk action to the Lagrangian one.

The AdS/CFT correspondence has been tested by computing various two-
and three-point functions of local operators in $D=4$, ${\cal N}=4$
supersymmetric Yang-Mills theory on $AdS_5$. 
In particular two-point functions corresponding to
scalars \cite{W} - \cite{Vol}, vectors 
\cite{W}, \cite{FMMR}, \cite{CNSS}, \cite{MV2}, spinors 
\cite{MV2} - \cite{GKPF}, the Rarita-Schwinger field
\cite{Cor} - \cite{MV4}, antisymmetric form fields 
\cite{AF3} - \cite{MR} and the graviton 
\cite{AF2}, \cite{LT}, \cite{MV3} were computed on the $AdS$ background. 
The only field in the supergravity spectrum, found in 
\cite{KRN}, that has evaded the attention is the massive
symmetric second-rank tensor field. In this paper we fill in the gap by 
computing the remaining two-point Green function.

We note that the Dirichlet boundary value problem for the massive symmetric
tensor field is nontrivial due to the fact that the equations of motion
for various components are coupled. 
Furthermore, in computing the two-point Green function we have to stick with
the regularization procedure described above in order to obtain the consistent
result. At the end of our computation we find that in the limit when the AdS 
mass vanishes the correlation function
reduces to that of the massless symmetric tensor field, i.e. the graviton.

\section{Equations of motion}
\setcounter{equation}{0}
The starting point in the calculation is the action for the symmetric
second-rank tensor $\phi_{\mu \nu}$ on $AdS_{d+1}$ 
\cite{BKP}\footnote{Note that the AdS background satisfies restriction
(10) of \cite{BKP} and also to simplify our calculations we set the
parameter $\xi$ of \cite{BKP} equal to 1.}:
\bea
\nonumber&&
S\,[\phi_{\mu \nu}] = \frac{1}{2 \kappa^2} \int_{AdS} d^{d+1}x \sqrt{|g|}
               \left( \frac14 \nabla_\lambda \phi \nabla^\lambda \phi
              -\frac14 \nabla_{\lambda} \phi_{\mu \nu} 
	       \nabla^\lambda \phi^{\mu \nu}
	      -\frac12 \nabla^\mu \phi_\mu^\nu \nabla_\nu \phi 
	      +\frac12 \nabla_\mu \phi^{\mu \nu} \nabla^\lambda 
\right.
\\\la{action}&&\hphantom{S\,[\phi_{\mu \nu}] = \frac{1}{2 \kappa^2} 
                         \int_{AdS} d^{d+1}x \sqrt{|g|} \left( \right.}  
    \left.    
    	       +\phi_{\lambda \nu} - 
	       \frac{d}{2} \phi_{\mu \nu} \phi^{\mu \nu}
	      +\frac{d}{4}\phi^2
              -\frac14 m^2 \phi_{\mu \nu} \phi^{\mu \nu} 
	      -\frac14 m^2 \phi^2 \right),
\eea
where $g$ is the determinant of the $AdS$ metric $g_{\mu \nu}$.
The action (\ref{action}) leads to the following equations 
of motion (\cite{KRN}, \cite{BKP})
\bea
\la{eom}
&&
\nabla^\lambda \nabla_\lambda \phi^\mu_\nu + (2-m^2) \phi^\mu_\nu = 0,
\\
\la{eom1}
&&
\phi^\mu_\mu =0, \qquad \nabla_\mu \phi^\mu_\nu = 0 .
\eea
The massive terms in (\ref{action}) destroy
the standard symmetry: $\d \phi_{\mu \nu} = \nabla_\mu \xi_\nu + \nabla_\nu 
\xi_\mu$,
which is present in the case of the massless symmetric tensor field.
As a result of this symmetry breaking one can no longer perform gauge fixing. 
Since $\phi^\mu_\nu$ is traceless, we can eliminate the component $\phi^0_0$ 
from the equations of motion by using the constraint $\phi^0_0 + \phi^i_i = 0$.
Let us introduce a concise notation: 
\bea
\nonumber
\phi = \phi^i_i , \quad \vphi^i = \phi^i_0 , \quad \vphi = \p_i \vphi^i.
\eea
Then starting from (\ref{eom}),(\ref{eom1}) one can obtain the
following system of coupled differential equations 
\bea
\la{simple}
&&
\vphi = \left( \p_0 - \frac{d+1}{x_0} \right) \phi,
\\
\la{2_diff}
&&
\left(\p_0 - \frac{d+1}{x_0} \right) \vphi^i + \p^k \phi^i_k = 0,
\\
\la{phi}
&&
\left( \p_0^2 - \frac{d+3}{x_0} \p_0 + \frac{2(d+2)}{x_0^2} + \Box 
- \frac{m^2}{x_0^2} \right) \phi = 0 .
\eea
Here eq.(\ref{phi}) corresponds to 
$\mu = \nu = 0$ in (\ref{eom}) while eqs.(\ref{simple}) and (\ref{2_diff}) 
follow from (\ref{eom1}). To simplify the notation, we chose the convention in 
which indices are raised and lowered with the flat metric 
$\delta^{\mu \nu} = x_0^{-2} g^{\mu \nu}$ so that, in particular, 
$\Box = \delta^{\mu \nu} \p_\mu \p_\nu $. 
The Fourier mode for the field $\phi$ vanishing at $x_0 \to \infty$ is given by:
\bea
\la{phi_solv}
&&
\phi(x_0, \k) = A(\eps, \k) x_0^{\frac{d}{2} + 2} K_\Delta( k x_0).
\eea
Here $k = | \k |$, 
$$
\Delta = \sqrt{\left( \frac{d}{2} \right)^2 + m^2}, 
$$
and $K_\Delta$ is a modified Bessel functions \cite{GR}:
\bea
\nonumber
K_\Delta(z) = \frac{\pi}{2} \frac{I_{-\Delta}(z) - I_\Delta(z)}{\sin \Delta \pi} \, ,
\qquad I_\Delta(z) = \sum_{k=0}^{\infty} \frac{1}{k! \Gamma(\Delta+k+1)} { \left(
\frac{z}{2} \right) }^{\Delta + 2k}.
\eea
Recall that the modified Bessel function satisfies the recurrence 
relations \cite{GR}:
\bea
\la{Mack}
K_{\Delta+1}(z) - K_{\Delta-1}(z) = \frac{2 \Delta}{z} K_\Delta(z) &,&
K_{\Delta+1}(z) + K_{\Delta-1}(z) = - 2 \frac{d}{dz} K_\Delta(z).
\eea
By differentiating (\ref{2_diff}) with respect to $x_i$ and using 
(\ref{simple}) and (\ref{phi}) one gets:
\bea
\la{rho'} 
&&
\p_i \p^j \phi^i_j = \left( \frac{d-1}{x_0} \p_0 + \Box - 
\frac{(d-1)(d+2) + m^2}{x_0^2} \right) \phi .
\eea
From (\ref{phi_solv}) and (\ref{rho'}) we can easily deduce the Fourier mode 
solution for the field
\bea
\nonumber
&&
\pi(x_0,\k) \equiv k_i k^j \phi^i_j (x_0,\k) =
\\
\nonumber
&& =
A(\eps, \k) \, x_0^{\frac{d}{2}} \,
\Biggl( (d-1)(x_0 k) K_{\Delta +1}(x_0 k) + 
\left( \Delta_- \left( \Delta_- - 1\right) + 
(x_0 k)^2 \right) K_\Delta(x_0 k) \Biggr),
\eea
where recurrence formulae (\ref{Mack}) were used. Here we introduced a concise
notation
$$
\Delta_- = \frac{d}{2} - \Delta, 
\qquad \Delta_+ = \frac{d}{2} + \Delta.
$$
Taking the ratio of $\phi$ and $\pi$ at $x_0 = \eps$ results
in the following relation
\bea
\nonumber
\pi (\k) &=& \eps^{-2} \,
\frac{(d-1)(\eps k) K_{\Delta+1}(\eps k) + 
\left( \Delta_- (\Delta_- - 1) + (\eps k)^2 \right)
K_\Delta(\eps k)}{K_\Delta(\eps k)} \, \phi(\k),
\eea
where $\pi (\k) = \pi (\eps, \k)$ and, similarly for other fields.
In the limit $ \eps \to 0$,
$\pi (\k) \sim \eps^{-2} \phi(\k)$ and therefore 
if we keep $\pi(\k)$ finite as $\eps \to 0$, then $\phi(\k)$ 
will tend to zero. On the other hand, keeping $\phi(\k)$ finite leads to the
divergence of $\pi(\k)$. Consequently, we ought to fix $\pi$ at $x_0 = \eps$.
Thus we have:
\bea
\nonumber
&&
\phi(x_0, \k) = \frac{x_0^{\frac{d}{2} + 2}}{\eps^{\frac{d}{2}}}
\frac{K_\Delta(x_0 k)}{(d-1)(\eps k) K_{\Delta+1}(\eps k) +
( \Delta_- (\Delta_- - 1) +(\eps k)^2)K_\Delta(\eps k)}
\pi (\k) ,
\\
\la{phi'}
\\
\nonumber
&&
\pi(x_0,\k) = 
\frac{x_0^{\frac{d}{2}} }{\eps^{\frac{d}{2}}}
 \frac{(d-1)(x_0 k) K_{\Delta +1}(x_0 k) + 
\left( \Delta_- (\Delta_- - 1) +(x_0 k)^2 \right) 
K_\Delta(x_0 k)}{(d-1)(\eps k) K_{\Delta +1}(\eps k) + \left( 
 \Delta_- (\Delta_- - 1) + (\eps k)^2 \right) K_\Delta(\eps k)} 
\pi(\k). 
\\
\la{rho}
\eea
The Fourier mode solution for the field $\vphi$ is found by
substituting (\ref{phi'}) into (\ref{simple}):
\bea
\la{solv_eta}
\vphi(x_0, \k) = \frac{x_0^{\frac{d}{2} +1} }{\eps^{\frac{d}{2}} }
\frac{(1 - \Delta_-) K_\Delta(x_0 k) - (x_0 k)K_{\Delta+1}(x_0 k) }{ 
(d-1)(\eps k) K_{\Delta+1}(\eps k) +
\left( \Delta_- (\Delta_- - 1) + (\eps k)^2 \right) 
K_\Delta(\eps k))} 
\pi (\k),
\eea
where once again recurrence formulae (\ref{Mack}) were used.
Next we turn our attention to the field $\vphi_i$. Here we need to use
equation
\bea
\la{2}
&&
\left( \Box -\frac{m^2}{x_0^2} \right) \vphi^i + \frac{2}{x_0} \p^i \phi = 
\p_0 \p^k \phi^i_k ,
\eea
which follows from $(i0)$-component of (\ref{eom}) by
taking into account (\ref{2_diff}).
Now we decompose $\vphi_i$ into the transversal and longitudinal parts:
$$
{\vphi}^{\perp}_i = \vphi_i - \frac{\p_i}{\Box} \vphi .
$$
Rewriting (\ref{2_diff}) and (\ref{2}) for the transversal part of $\vphi_i$ 
leads to:
\bea
\la{2_tr}
\left( \Box - \frac{m^2}{x_0^2} \right) {\vphi}^{\perp}_i &=& \p_0 \left(
\p_k \phi_i^k - \frac{\p_i}{\Box} \p_l \p^m \phi^l_m \right),
\\
\la{relation}
\left( \p_0 - \frac{d+1}{x_0} \right) {\vphi}^{\perp}_i &=&
- \left( \p_k \phi^k_i  - \frac{\p_i}{\Box} \p_l \p^m \phi^l_m \right) .
\eea
By differentiating (\ref{relation}) with respect to $x_0$ and taking into
account (\ref{2_tr}) we obtain an homogeneous differential equation for the 
field ${\vphi}^{\perp}_i$:
\be
\la{2_final}
\left( \p_0^2 - \frac{d+1}{x_0} \p_0 + \frac{d+1}{x_0^2} + \Box -
\frac{m^2}{x_0^2} \right)
{\vphi}^{\perp}_i = 0.
\ee
The Fourier mode solution is given by:
\bea
\la{ttemp}
&&
{\vphi}^{\perp}_i (x_0,\k) = B_i(\eps, \k) \, x_0^{\frac{d}{2} +1} \, 
K_\Delta(x_0 k) .
\eea
To find $B_i(\eps, \k)$ substitute (\ref{ttemp}) into (\ref{relation})
to obtain the following formula
\be
\la{rel}
k_l \phi^l_i(x_0,\k) - \frac{k_i}{k^2} \pi(x_0, \k) =
i B_i (\eps, \k) \, x_0^{\frac{d}{2}} \, \left(
(x_0 k) K_{\Delta+1}(x_0 k) + \Delta_- K_\Delta(x_0 k) \right).
\ee
Taking the ratio of (\ref{rel}) and (\ref{ttemp}) at $x_0 = \eps$ we find:
\bea
\nonumber
k^l \phi_l^i(\k) - \frac{k^i}{k^2} \pi (\k) &=&
i \eps^{-1} \, \frac{
(\eps k) K_{\Delta+1}(\eps k)}{ K_\Delta(\eps k)
+ \Delta_- K_\Delta(\eps k) } \, {\vphi}^{\perp}_i(\k) .
\eea
Using the same arguments as before one finds that in order to avoid the 
divergence at $\eps \to 0$ the solutions for the field ${\vphi}^{\perp}_i$ 
and $k_l \phi^l_i - \frac{k_i}{k^2} \pi$ should take the following 
form
\bea
\la{tvphi}
&&
{\vphi}^{\perp}_i (x_0, \k) = 
-i \eps 
\frac{x_0^{\frac{d}{2} +1}}{\eps^{\frac{d}{2}}}
\frac{K_\Delta(x_0 k)}{
(\eps k)K_{\Delta+1}(\eps k)
+ \Delta_- K_\Delta(\eps k)}
\left( k^l \phi_l^i(\k) - 
\frac{k^i}{k^2} \pi (\k) \right) ,
\\
\nonumber
&&
k^l \phi_l^i (x_0,\k) - \frac{k^i}{k^2} \pi (x_0,\k) =
\frac{x_0^{\frac{d}{2}}}{\eps^{\frac{d}{2}}}  \frac{
(x_0 k) K_{\Delta+1}(x_0 k) + \Delta_- K_\Delta(x_0 k)}{
(\eps k) K_{\Delta+1}(\eps k)
+ \Delta_- K_\Delta(\eps k)} 
\left( k^l \phi_l^i (\k) - 
\frac{k^i}{k^2} \pi (\k) \right) .
\\
\la{phi_i}
\eea
Setting $\mu = i$, $\nu= j$ in (\ref{eom}) and taking into
account (\ref{eom1}) we arrive at the following equation
\be
\la{3_simple}
\left( \p_0^2 - \frac{d-1}{x_0} \p_0 + \Box - \frac{m^2}{x_0^2} \right) 
\phi^i_j = \frac{2}{x_0} \left( \p_j \vphi^i + \p^i \vphi_j \right) + 
\frac{2 \d^i_j}{x_0^2} \phi.
\ee
Let us introduce the transversal traceless part of $\phi^i_j$:
\be
\la{def}
{\ttphi}^i_j = 
\phi^i_j - \frac{1}{\Box} \p^i \p_k \phi^k_j -\frac{1}{\Box}
\p_j \p^k \phi^i_k + \frac{1}{\Box^2} \p^i \p_j \p_k \p^m \phi^k_m
+
\frac{1}{d-1} \left( \frac{\p^i \p_j}{\Box} - \d^i_j\right) 
\left( \phi - \frac{\p_k \p^l}{\Box} \phi^k_l \right).
\ee
Rewriting (\ref{3_simple}) for the transversal traceless part yields:
\be
\la{3_final}
\left( \p_0^2 - \frac{d-1}{x_0} \p_0  + \Box - \frac{m^2}{x_0^2} \right)
{\ttphi}^i_j = 0 .
\ee
The Fourier mode solution of eq.(\ref{3_final}) is 
\bea
\la{tt}
&&
{\ttphi}^i_j (x_0, \k) =
\frac{x_0^{\frac{d}{2}}}{\eps^{\frac{d}{2}}}
\frac{K_\Delta(x_0 k)}{K_\Delta(\eps k)} {\ttphi}^i_j (\k).
\eea
Taking into account (\ref{def}), (\ref{tt}), (\ref{phi'}),
(\ref{rho}), and (\ref{phi_i}) we obtain:
\bea
\nonumber&&
\phi^i_j (x_0,\k) =
\frac{ x_0^{ \Delta_- }}{ \eps^{\Delta_-} }\Biggl[
\frac{\K_\Delta(x_0 k)}{\K_\Delta(\eps k)} {\ttphi}^i_j(\k)
+
\frac{ \K_{\Delta+1}(x_0 k) + \Delta_- \K_{\Delta}(x_0 k)}{
       \K_{\Delta+1}(\eps k) + \Delta_- \K_{\Delta}(\eps k)}  
\left(
\frac{k^i k_l}{k^2} \phi^l_j(\k) + \frac{k_j k^l}{k^2} \phi^i_l(\k) 
\right.
\\\nonumber&&\hphantom{\phi^i_j (x_0,\k) =
\frac{ x_0^{ \Delta_- }}{ \eps^{\Delta_-} }\Biggl[} 
\left.
-2 \frac{k^i k_j}{k^4} \pi(\k) \right) 
+
\frac{ (d-1) \K_\Delta(x_0 k) + \left( \Delta_- (\Delta_- - 1) + (x_0 k)^2 
\right) \K_\Delta(x_0 k)}{
      (d-1) \K_\Delta(\eps k) + \left( \Delta_- (\Delta_- - 1) + (\eps k)^2 
\right) \K_\Delta(\eps k)} \times
\\\nonumber&&\hphantom{\phi^i_j (x_0,\k) =
\frac{ x_0^{ \Delta_- }}{ \eps^{\Delta_-} }\Biggl[} 
\times \frac{k^i k_j}{k^4} \pi (\k)
-
\frac{1}{d-1} \left( \frac{k^i k_j}{k^2} -\d^i_j \right) \times
\\\nonumber&&\hphantom{\phi^i_j (x_0,\k) =
\frac{ x_0^{ \Delta_- }}{ \eps^{\Delta_-} }\Biggl[} 
\times \left(
\frac{ (x_0 k)^2 \K_\Delta(x_0 k)}{(d-1)(\eps k) \K_{\Delta+1}(\eps k) +
(\Delta_- (\Delta_- - 1) + (\eps k)^2) \K_\Delta(\eps k)} \right.
\\\la{final1}&&\hphantom{\phi^i_j (x_0,\k) =
\frac{ x_0^{ \Delta_- }}{ \eps^{\Delta_-} }\Biggl[} 
\left. -
\frac{ (d-1) \K_\Delta(x_0 k) + \left( \Delta_- (\Delta_- - 1) + (x_0 k)^2 
\right) \K_\Delta(x_0 k)}{
       (d-1) \K_\Delta(\eps k) + \left( \Delta_- (\Delta_- - 1) + (\eps k)^2 
\right) \K_\Delta(\eps k)}
\right) \frac{\pi(\k)}{k^2} \Biggr],
\eea
while taking into account (\ref{tvphi}) and (\ref{solv_eta}) gives:
\bea
\nonumber&&
\vphi_i (x_0 , \k) =
-(i \eps) {\left( \frac{x_0}{\eps} \right)}^{1 + \Delta_-}
\Biggl[
\frac{\K_\Delta(x_0 k)}{\K_{\Delta+1}(\eps k) + \Delta_- \K_\Delta(\eps k)} 
\left( k_l \phi^l_i (\k) - \frac{k_i}{k^2} \pi (\k) \right)
\\\nonumber&&\hphantom{\vphi_i (x_0 , \k) =
-(i \eps) {\left( \frac{x_0}{\eps} \right)}^{1 + \Delta_-}}  
-
\frac{ (1 - \Delta_- ) \K_\Delta(x_0 k) - \K_{\Delta+1}(x_0k)}{
(d-1)\K_{\Delta+1}(\eps k) + (\Delta_- (\Delta_- - 1) + (\eps k)^2) 
\K_\Delta(\eps k)} \frac{k_i}{k^2} \pi(\k)
\Biggr] ,
\\\la{final2}\eea
where we introduced a concise notation: $\K_\Delta(z) = z^\Delta K_\Delta(z)$. 

\section{Two-point Green function}
\setcounter{equation}{0}

To compute the Green function in the framework of the $AdS/CFT$ correspondence
we need to evaluate the on-shell value of the action. Taking into
account equations of motion (\ref{eom}) and (\ref{eom1}) one finds that the 
on-shell value of (\ref{action}) is
\bea
\nonumber
&&
S_{on-shell} =
\frac{\eps^{-d+1}}{8 \kappa^2} \int_{x_0 =\eps} d^d x \left( \phi^i_j \p_0 \phi^j_i
-\phi \p_0 \phi + 2 \phi \vphi - 2 \vphi^k \p_i \phi^i_k +
\frac{2(d+1)}{\eps} \left( \phi^2 + \vphi^k \vphi_k \right) \right) .
\la{actionI}
\eea
Let us first consider the contribution to $S_{on-shell}$ that depends
locally on boundary fields, i.e. does not contain the normal derivative
$\p_0$. We expect that such terms do not contribute to the non-local part of
$S_{on-shell}$. So we need to consider the behavior of $\phi$ and $\vphi^k$
on the boundary of $AdS$. To this end we note that according to (\ref{phi'})
the field $\phi(\k)$ is local since
$$
\frac{K_\Delta(\eps k)}{ (d-1) (\eps k) K_{\Delta+1}(\eps k) +
\left( \Delta_- (\Delta_- - 1) + (\eps k)^2 \right) K_\Delta(\eps k)} =
\frac{1}{\Delta_+ (\Delta_+ - 1)} + O (\eps^2 k^2),
$$
while expanding solution (\ref{final2}) in $\eps$ gives:
\bea
\la{eta_loc}
&&
\vphi^i(\k) =
-(i \eps) \Biggl[ \frac{1}{\Delta_+} 
\left( k^l \phi^i_l(\k) - \frac{k^i}{k^2} \pi (\k) \right)
+
\frac{1}{\Delta_+} \frac{k^i}{k^2} \pi (\k) 
\Biggr] + \mbox{local}.
\eea
Here we included only non-local terms. In deriving
(\ref{eta_loc}) use was made of the power series expansion of the modified
Bessel function \cite{GR}:
\be
\la{Bessel}
\K_\Delta(z) = 2^{\Delta-1} \Gamma(\Delta) \left( 1 + \frac{z^2}{4 (1-\Delta)}
+ \ldots \right),
\ee
where ellipsis indicate terms of order $z^4$ and higher
which evidently lead to local expressions as well as terms of order 
$z^{2 \Delta + 2}$ which will become negligible in the limit $\eps \to 0$.
From expression (\ref{eta_loc}) it follows that
$\vphi_i(\k)$ is a local field. Consequently, the terms in (\ref{actionI}) that 
depend only on the value of fields at the boundary do not contribute to 
the non-local part of $S_{on-shell}$ as we expected.

Next we consider terms with the normal derivative $\p_0$.
In evaluating such terms it is useful to employ the following identity
\be
\la{identity}
\frac{d}{d x_0} \left( { \left( \frac{x_0}{\eps} \right) }^{\gamma}
\frac{F(x_0 k)}{F(\eps k)} \right|_{x_0 = \eps}  =
\frac{\gamma}{\eps} + \frac{k}{\eps} \frac{d}{dk} \ln F (k \eps) .
\ee
Taking into account solution (\ref{phi'}), 
identity (\ref{identity}) and expansion (\ref{Bessel}) we find that
$\p_0 \phi (\eps, \x)$ is equal to
\bea
\nonumber
&&
\p_0 \phi (\eps, \x) =
\int \frac{d^d k}{(2 \pi)^d} e^{-i \k \x}
 \left( \frac{2 + \Delta_-}{\eps} + 
O (\eps k^2)
 \right) \eps^2
\left( \frac{1}{ \Delta_+ (\Delta_+ - 1)} + O(\eps^2 k^2) \right) 
\pi (\k).
\eea
Clearly $\p_0 \phi$ is a local expression and therefore does not
contribute to the non-local part of $S_{on-shell}$.
Consequently, the non-local part is entirely determined by the
following expression
\bea
\la{actionII}
&&
\frac{\eps^{-d+1}}{8 \kappa^2} \int_{x_0 = \eps} d^d x \,
 \phi^j_i (\x) \int \frac{d^d k}{(2 \pi)^d} e^{-i \k \x} \Bigl( \p_0
\phi^i_j (x_0, \k) \Bigr) .
\eea
Taking into account solution (\ref{final1}),
identity  (\ref{identity}) and expansion (\ref{Bessel}) we see that
the expression for $\p_0 \phi^i_j$ gets three different contributions:
the first contribution comes from differentiating the ratio
$\frac{x_0}{\eps}$ raised to the power $ \Delta_- $ or 
$\left( 2 + \Delta_- \right)$; the second
contribution comes from $O (\eps k^2)$ terms in the power expansion of the
logarithmic derivative of various functions in (\ref{final1}) and the third
contribution comes from $O( \eps^{2\Delta -1} k^{2\Delta})$ terms in the same
expansion.
From (\ref{final1}) we can easily read off the first contribution which is
equal to
\bea
\nonumber
&&
 \frac{\Delta_-}{\eps} {\ttphi}^i_j(\k) +
 \frac{\Delta_-}{\eps}
  \left( \frac{k^i k_l}{k^2} \phi^l_j(\k) + \frac{k_j k^l}{k^2} \phi^i_l(\k)
-2\frac{k^i k_j}{k^4} \pi (\k) \right) 
+ \frac{\Delta_-}{\eps} \frac{k^i k_j}{k^4} \pi (\k)
\\
\nonumber
&&
-\frac{1}{d-1} \left( \frac{k^i k_j}{k^2} - \d^i_j \right)
\left( \frac{2 + \Delta_- }{\eps} (\eps k)^2  
\frac{1}{\Delta_+ (\Delta_+ -1) } -
\frac{\Delta_-}{\eps} \right) \frac{\pi (\k)}{k^2} 
\\
\la{s1}
&& 
=
-\frac{2}{d-1} \frac{k^i k_j}{k^2} \frac{\eps}{
\Delta_+ (\Delta_+ - 1)} \pi (\k) + \,\, \mbox{local} \,\, ,
\eea
where we took into account the expansion of $\phi^i_j$ into the 
transversal and longitudinal parts given by 
(\ref{def}).
Next, $O(\eps k^2)$ terms from the power series expansion of the logarithmic 
derivative give:
\bea
\nonumber&&
\frac{\eps k^2}{2} \Biggl[
\frac{1}{1 - \Delta} {\ttphi}^i_j (\k) +
\frac{\Delta_+ - 2}{(1 - \Delta)( \Delta_+ )}
\left( \frac{k^i k_l}{k^2} \phi^l_j(\k) + \frac{k_j k^l}{k^2} \phi^i_l(\k)
- 2 \frac{k^i k_j}{k^4} \pi (\k) \right) 
\\\nonumber&&\hphantom{\frac{\eps k^2}{2} }
+\frac{(\Delta_+ - 2)(\Delta_+ - 3)}{(1 - \Delta)
\Delta_+ (\Delta_+ - 1)} \frac{k^i k_j}{k^4} \pi (\k)
-\frac{1}{d-1} \left( \frac{k^i k_j}{k^2} -\d^i_j \right)
\left( 
\frac{(\eps k)^2}{(1-\Delta)\Delta_+ (\Delta_+ - 1)} 
\right.
\\\nonumber&&\hphantom{\frac{\eps k^2}{2} }
\left.
- 
\frac{(\Delta_+ - 3)(\Delta_+ - 2)}{(1-\Delta) \Delta_+ (\Delta_+ - 1)} \right) 
\frac{\pi(\k)}{k^2} \Biggr]
\\\la{lowest}&&
=
\frac{2}{d-1} \frac{k^i k_j}{k^2} \frac{\eps}{\Delta_+ (\Delta_+ - 1)} 
\pi (\k) + \mbox{local},
\eea
where we used (\ref{Bessel}) and (\ref{def}).
Adding together (\ref{lowest}) and (\ref{s1}), the non-local part cancels
leading to a purely local expression which does not contribute to the non-local
part of $S_{on-shell}$.
Finally, $O( \eps^{2\Delta-1} k^{2 \Delta})$ terms in the expansion
of the logarithmic derivative give:
\bea
\nonumber&&\hphantom{=}
\eta \, \eps^{2 \Delta - 1} k^{2 \Delta} \Biggl[
{\ttphi}^i_j(\k)
+ \frac{\Delta_-}{\Delta_+} \left( \frac{k^i k_l}{k^2} \phi^l_j(\k) + 
\frac{k_j k^l}{k^2} \phi^i_l(\k)
- 2 \frac{k^i k_j}{k^4} \pi (\k) \right) +
\frac{ \Delta_- (\Delta_- - 1)}{
\Delta_+ (\Delta_+ - 1)} \frac{k^i k_j}{k^4} \pi (\k)
\\\nonumber&&\hphantom{=\eta \, \eps^{2 \Delta - 1} k^{2 \Delta} \Biggl[}
+
\frac{1}{d-1} \left( \frac{k^i k_j}{k^2} - \d^i_j \right)
\frac{ \Delta_- (\Delta_- - 1)}{
\Delta_+ (\Delta_+ - 1)} 
\frac{\pi (\k)}{k^2} \Biggr] 
\\\la{actionV}&&
=\eta \, \eps^{2\Delta -1} k^{2 \Delta}
\Biggl[ \phi^i_j(\k) 
- 
\frac{2\Delta}{ \Delta_+ } 
\left( \frac{k^i k_l}{k^2} \phi^l_j(\k) + \right.
\left. +
\frac{k_j k^l}{k^2} \phi^i_l(\k) 
\right) - \frac{4 \Delta (1 - \Delta)}{ \Delta_+ (\Delta_+ - 1) }
\frac{k^i k_j}{k^4} \pi (\k) \Biggr] .
\eea
Here,
$$
\eta = - 
\frac{ \Delta }{2^{2\Delta - 1}} \frac{\Gamma(1 - \Delta)}{\Gamma (1 + \Delta)}.
$$
In the process of deriving (\ref{actionV}) we dropped all terms 
containing $\d^i_j$ since such terms are negligible in the limit
$\eps \to 0$. To understand why this is so, note that
when $\d^i_j$ is contracted with $\phi^j_i(\x)$ from
(\ref{actionII}) it gives the trace of $\phi^i_j(\x)$ which
according to (\ref{phi'}) is order $O(\eps^2)$ at $x_0 = \eps$. 

Now putting together (\ref{actionII}) and (\ref{actionV}) we arrive at the
following formula for $S_{on-shell}$
\bea
\nonumber
&&
S_{on-shell} [\phi^i_j] =
\frac{\eta \eps^{2\Delta - d}}{8 \kappa^2} \int \int d^d x \, d^d y \,
{\bf \Phi}^j_i (\x) \, {\bf \Phi}^r_s (\y)
\int \frac{d^d k}{(2 \pi)^d} e^{-i \k (\x - \y) } \, k^{2 \Delta} 
\times
\\\nonumber
&&\hphantom{S_{on-shell} [\phi^i_j] =}
\times \left[ \frac12 \d^i_r \d^s_j + \frac12 \d^{is} \d_{jr} 
- 
\frac{\Delta}{\Delta_+} \left( \frac{k^i k_r}{k^2} \d_j^s + 
\frac{k^i k^s}{k^2} \d_{jr} + 
\frac{k_j k^s}{k^2} \d_r^i + 
\frac{k_j k_r}{k^2} \d^{is} \right) \right.
\\\nonumber&&\hphantom{S_{on-shell} [\phi^i_j] = \times}
+ \frac{2 \Delta}{\Delta_+ (\Delta_+ - 1)} 
  \left( 
   2(\Delta-1) \frac{k^i k_j k_r k^s}{k^4} 
   + \frac{k_r k^s}{k^2} \d^i_j + \frac{k^i k_j}{k^2} \d^s_r  
   \right.
\\\la{act}&&\hphantom{S_{on-shell} [\phi^i_j] = \times + 
                      \frac{2 \Delta}{\Delta_+ (\Delta_+ - 1)} \left( \right.}
\left.
-\frac{ \Delta_+^2 - \Delta_-}{2 d \Delta } \d^i_j \d^s_r 
\right) \Biggr],
\eea
where we introduced the traceless part of $\phi^i_j$:
$$
{\bf \Phi}^i_j(\x) = \phi^i_j(\x) - \frac{\d^i_j}{d} \phi(\x) .
$$
In order to complete the calculation of the two-point Green function we need to
evaluate the integral over $\k$ in (\ref{act}). To this end, we employ
the standard formula for the Fourier transformation of generalized functions
\cite{GSh}:
\bea
\la{gelfand}
&&
\int \frac{d^d k}{(2 \pi)^d} e^{- i \k \x} \frac{ k^{i_1} k^{i_2} \cdots
k^{i_n} }{ k^n} {| \k |}^{\d} =
\frac{2^{\d-n}  \Gamma \left( \frac{d+\d-n}{2} \right)}{
	       \pi^{d/2} \Gamma \left( \frac{-\d+n}{2} \right)}
{(i)}^n \p_{i_1} \p_{i_2} \cdots \p_{i_n}
\frac{1}{{| \x |}^{d+\d-n}}.
\eea
With the help of (\ref{gelfand}) we find that the non-local part of the
on shell value of the action is equal to:
\bea
\nonumber&&
S_{on-shell}[\phi^i_j] =
\frac{C_{d, \Delta} \eps^{2\Delta-d}}{2}
\int \int d^d x \, d^d y \frac{ {\bf \Phi}^j_i(\x)
{\bf \Phi}^r_s(\y)}{
{|\x - \y|}^{2 \Delta +d}} \Biggl[ \frac12 
J^i_r(\x -\y) J^s_j(\x-\y)  
\\\nonumber&&\hphantom{S_{on-shell}[\phi^i_j] =
\frac{C_{d, \Delta} \eps^{2\Delta-d}}{2}
\int \int d^d x \, d^d y \frac{ {\bf \Phi}^j_i(\x)
{\bf \Phi}^r_s(\y)}{
{|\x - \y|}^{2 \Delta +d}}} 
+ \frac12 J^{is} (\x -\y) J_{jr}(\x-\y) 
- \frac{1}{d} \d^i_j \d^r_s \Biggr],
\eea
where we introduced
\bea
\la{constant}
&&
J^i_j(\x) = \d^i_j - 2 \frac{x^i x_j}{| \x |^2} \qquad \mbox{and} \qquad
C_{d,\Delta} = \frac{ \Delta (\Delta_+ + 1) \Gamma(\Delta_+ ) }{
2 \pi^{\frac{d}{2}} \kappa^2 (\Delta_+ - 1) \Gamma(\Delta) } 
.
\eea
From this we can easily deduce the two-point function of local operators
in the boundary CFT corresponding to the massive symmetric traceless rank two
tensor field ${\bf \Phi}^i_j$:
\bea
\nonumber
< {\cal O}^i_j (\x) {\cal O}^s_r (\y) > =
\frac{C_{d,\Delta}}{{|\x - \y|}^{2 \Delta+d}} 
\Biggl[ \frac12 J^i_r (\x -\y) J_j^s (\x-\y) + \frac12 J^{is} (\x -\y) 
J_{jr}(\x-\y) -\frac{1}{d} \d^i_j \d_r^s \Biggr] .
\eea
Here we performed rescaling in order to remove the regularization parameter
$\eps$. Note that in the limit $m^2 \to 0$ or, equivalently, $\Delta \to 
\frac{d}{2}$,
the obtained expression correctly reproduces the two-point function
corresponding to the massless symmetric tensor field (graviton) 
\cite{AF2}, \cite{LT}, \cite{MV3}.

{\bf ACKNOWLEDGEMENT}

\noindent The author would like to thank S.Frolov and G.Arutyunov for 
valuable discussions and A.Slavnov for the support during the preparation 
of the manuscript.


\end{document}